\documentclass[aps,prb,twocolumn,floatfix,amsmath,amssymb]{revtex4}
\usepackage[final]{graphicx}
\usepackage{bm}
\usepackage{afterpage}
\usepackage{color}
\usepackage{comment}
\usepackage[colorlinks=true,urlcolor=blue,linkcolor=blue,citecolor=blue]{hyperref}

\emergencystretch=\maxdimen
\begin{document}

\title{Antiferromagnetic transitions of Dirac fermions in three dimensions}

\author{Yiqun Huang}
\affiliation{Department of Physics, Beijing Normal University, Beijing, 100875, China}

\author{Huaiming Guo}
\email{hmguo@buaa.edu.cn}
\affiliation{Department of Physics, Key Laboratory of Micro-Nano Measurement-Manipulation and Physics (Ministry of Education), Beihang University,
Beijing, 100191, China}

\author{Joseph Maciejko}
\affiliation{Department of Physics and Theoretical Physics Institute, University of Alberta, Edmonton, Alberta, Canada T6G 2E1}

\author{Richard T. Scalettar}
\affiliation{Physics Department, University of California, Davis, CA
95616, USA}

\author{Shiping Feng}
\affiliation{Department of Physics, Beijing Normal University, Beijing, 100875, China }

\begin{abstract}
We use determinant quantum Monte Carlo (DQMC) simulations to study the role of electron-electron interactions on three-dimensional (3D) Dirac fermions based on the $\pi$-flux model on a cubic lattice. We show that the Hubbard interaction drives the 3D Dirac semimetal to an antiferromagnetic (AF) insulator only above a finite critical interaction strength and the long-ranged AF order persists up to a finite temperature. We evaluate the critical interaction strength and temperatures using finite-size scaling of the spin structure factor. The critical behaviors are consistent with the (3+1)d Gross-Neveu universality class for the quantum critical point and 3D Heisenberg universality class for the thermal phase transitions. We further investigate correlation effects in birefringent Dirac fermion system. It is found that the critical interaction strength $U_c$ is decreased by reducing the velocity of the Dirac cone, quantifying the effect of velocity on the critical interaction strength in 3D Dirac fermion systems. Our findings unambiguously uncover correlation effects in 3D Dirac fermions, and may be observed using ultracold atoms in an optical lattice.
\end{abstract}


\maketitle

\section{Introduction}
Electrons propagating on the honeycomb lattice have a linear energy-momentum dispersion, analogous to that of the two-dimensional Dirac equation\cite{castro2009,novoselov2007}. This novel state is called the Dirac semimetal, and has attracted great interest.
One aspect of the many studies is the interaction-driven quantum phase transition between the semimetal and various ordered phases.
While the semimetal is robust to weak interactions due to the vanishing density of states, AF long-ranged order develops for strong Hubbard interactions\cite{sorella1992}. There is no intermediate unconventional phase, such as a quantum spin liquid, between the semimetal and AF Mott insulator, and the phase transition is a direct and continuous one\cite{herbut2006,meng2010,sorella2012,assaad2013,sorella2016}. The physics is made even richer by the interaction-generated topological states in the extended Hubbard model\cite{raghu2008}. Although the mean-field theory predicts a quantum anomalous Hall effect and a quantum spin Hall effect stabilized by the nearest and next-nearest neighbor interactions, unbiased numerical methods find no evidence of their existence and support trivial ordered phases\cite{Capponi2016,Rachel2018}. Instead, the intriguing interaction-driven topological mechanism takes effect for a quadratic dispersion on the kagome and checkerboard lattices\cite{wuhanqing2016,zhuwei2016}. Another equally interesting quantum phase is topological superconductivity, which is believed to arise in the doped Hubbard model on a honeycomb lattice\cite{nandkishore2012,Black_Schaffer_2014}.

The quantum criticality of Dirac fermions has been extensively studied in recent literature. The critical behavior between the semimetal and various ordered phases is strongly affected by the gapless fermionic excitations, giving rise to a fermionic quantum critical point. The low-energy effective theory is the celebrated Gross-Neveu theory, which contains both a bosonic order parameter and Dirac fermions coupled by Yukawa-like terms\cite{herbut2006,herbut2009,herbut2009a,joseph2020}. Various universality classes are possible depending on the symmetry group of the order parameter and the number of fermion components. The $N=4$ chiral Ising class has been investigated in terms of spinless fermions with nearest-neighbor repulsion on the honeycomb and $\pi$-flux square lattices\cite{wang2014,li2015,wessel2016}. Charge-density-wave transitions in the spinful Holstein model are verified to be in the $N=8$ chiral Ising class\cite{zhangyuxi2019,chenchuang2019,feng2019,zhang2020}. The $N=8$ chiral Heisenberg criticality of the Hubbard model on the honeycomb and $\pi$-flux lattices has been studied by DQMC recently\cite{assaad2015,sorella2016,guo2018}. The transition between Dirac semimetal  and Kekul\'e valence-bond solid belongs to the chiral XY class, whose critical exponents have been calculated using renormalization-group analysis and DQMC simulations\cite{Lang_2013,zhou_2016,lizixiang2017,yaohong2017,otsuka_2018}.
Remarkably, the intriguing space-time supersymmetry, long sought in high-energy physics, has been found to emerge at the critical point between the semimetal and pair density wave\cite{yaohong2015}.

Considering the rich properties of 2D Dirac semimetals, it is natural to extend the study of correlation effects to 3D Dirac fermions. The idea receives a further boost from the remarkable progress in the field of 3D topological semimetals\cite{vishwanath2018}. Over the past few years, Dirac and Weyl fermions have been predicted and experimentally confirmed in a number of solid-state materials, prototypical examples including: TaAs, Cd$_{3}$As${_{2}}$, and Na$_{3}$Bi etc.\cite{zjwang2013,zkliu2014,zkliu2014b,hding2015,syxu2015} In Weyl fermions, while short-range interactions are perturbatively irrelevant, sufficiently strong interactions can induce a series of novel states, which can be axionic charge density wave\cite{wangzhong2013,bitan2015,bitan2017,gooth2019},antiferromagnetism\cite{zhai2016},spin density wave\cite{rachel2016}, chiral excitonic insulator\cite{aji2012}, and superconductivity with finite-momentum Fulde-Ferrell-Larkin-Ovchinnikov pairing\cite{moore2012,aji2013,maciejko2014,burkov2015,zhoutao2016}. Besides, novel correlation effects have been predicted such as non-Fermi-liquid and anisotropic Coulomb screening in anisotropic and multi-Weyl semimetals\cite{yang2014,hhlai2015,zhang2017,liu2017,jianshaokai2017,liu2018,moon2018,torres2018,han2019}.

Unlike Weyl fermions, 3D Dirac fermions are four-component complex spinors with the presence of both time-reversal and inversion symmetries. Since Dirac points have four-fold degeneracy, materials hosting them can be viewed as '3D graphene'. 3D Dirac semimetals can be realized at a quantum critical point in a normal-topological insulator transition\cite{Murakami_2007}. They can also appear from band inversion, or are enforced by symmetry\cite{vishwanath2018}. Although 3D Dirac semimetals have been extensively studied theoretically and experimentally, attempts to study correlation effects are rare\cite{gonzalez2015}.

In the manuscript, we study correlation effects in 3D Dirac fermions based on a toy model on the
cubic lattice, with $\pi$-flux through the faces (known as 3D $\pi$-flux model)\cite{hosur2010,mazzucchi2013,hayami2014,zhu2017}. For a specific choice of gauge, the model is only composed of positive and negative nearest-neighbor hopping terms. Unlike the usual 3D Dirac Hamiltonian, the current model is free of spin-orbit coupling, and thus can be numerically simulated using large-scale sign-problem-free quantum Monte Carlo method, providing a unique opportunity to exactly investigate the many-body physics in 3D Dirac fermions. In addition, 3D birefringent Dirac fermions can be realized by
modulating the hoppings, which provides a platform to study the dependence of the AF critical
interaction on the Fermi velocity\cite{kennett2011,kennett2014,guo2018}.

 The paper is organized as follows: Section 2 introduces the precise model we will investigate, along with our computational methodology. Section 3 presents the mean-field calculations. The order parameter is determined self-consistently, and the mean-field phase diagram is mapped out. Section 4 shows the DQMC results. We first evaluate the specific heat, whose evolution with the interaction clearly reflects the underlying semimetal-AF insulator transition. Then we calculate the equal-time spin structure factor for various lattice sizes and temperatures. The critical interaction strength and temperatures are determined using finite-size
scaling. We also present the result of 3D birefringent Dirac fermions to show the effect of the Fermi velocity on the critical interaction strength. Finally we offer concluding remarks in Section 5.


\section{The model and method}

\begin{figure}[htbp]
\centering \includegraphics[width=5.cm]{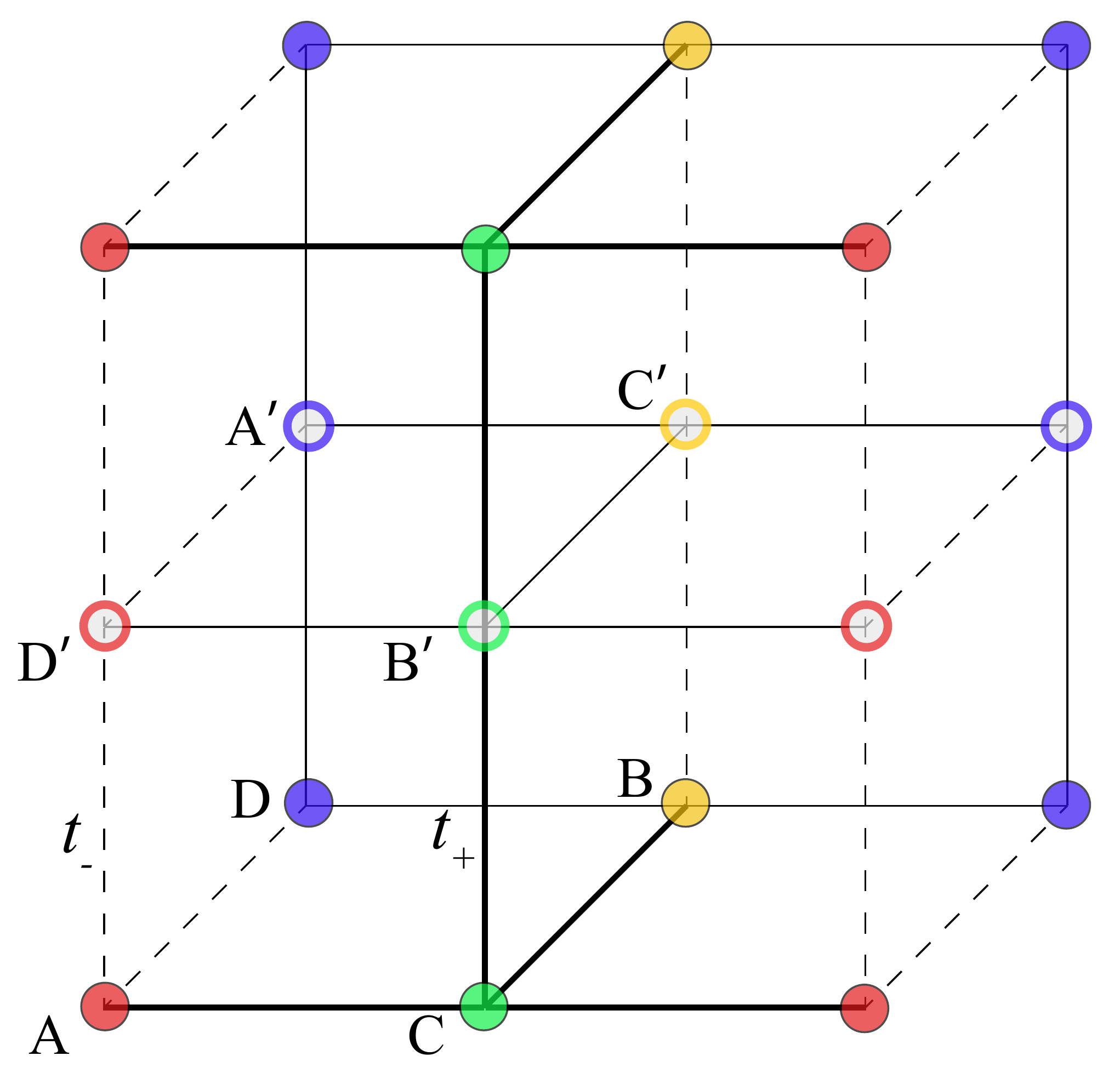} \caption{The cubic lattice with each face threaded by a $\pi$-flux. The solid (dashed) lines represent positive (negative) hoppings. The hopping amplitudes are parameterized as $t_{\pm}=(1\pm \alpha)t$ to generate two-species Dirac fermions. Here $t_{+}=1$ is set to fix the bandwidth. }
\label{fig1}
\end{figure}

\begin{figure}[htbp]
\centering \includegraphics[width=7.cm]{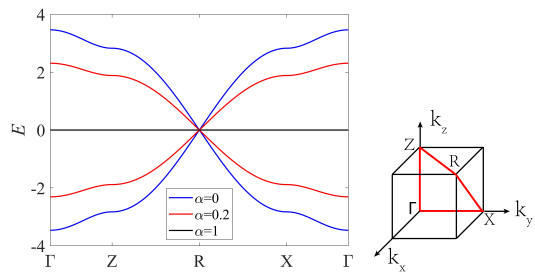} \caption{Energy spectrum of the $\pi$-flux cubic lattice for different values of $\alpha$. The blue bands are for $\alpha=0$. There are two branches for $\alpha\neq 0$. While the inner branch is the blue one which is the same for all $\alpha$, the outer one changes with $\alpha$, and becomes exactly flat in the $\alpha=1$ limit. The right figure shows the high-symmetry points in the Brillouin zone. Specifically $R$ represents the momentum point $(\pi/2,\pi/2,\pi/2)$.}
\label{fig2}
\end{figure}

We consider a Hamiltonian describing
3D Dirac fermions on a cubic lattice
where each plaquette is threaded with half a flux quantum,
$\frac{1}{2}\Phi_{0} =hc/(2e)$,
\begin{equation}\label{eq1}
H_0=\sum_{\langle lj \rangle \sigma}
t_{lj}e^{i\chi_{lj}}c^\dag_{j\sigma}c^{\phantom{\dag}}_{l\sigma},
\end{equation}
where $c^\dag_{j\sigma}$ and
$c^{\phantom{\dag}}_{j\sigma}$
are the creation and annihilation operators at site $j$
with spin $\sigma=\uparrow, \downarrow$.
The hopping amplitudes between the nearest-neighbor sites
$l$ and $j$ are $t_{lj}=t$, which we set to 1 as the unit of energy. $\chi_{lj}$ is the Peierls phase arising from
the magnetic flux
$\chi_{lj}=\frac{2\pi}{\Phi_{0}}\int_{{\bf x}_l}^{{\bf x}_j}
{\bf A}\cdot d{\bf x}$ with ${\bf A}$ the vector potential.
A particular gauge choice is shown in Fig.~1, where the solid (dashed) line represents hopping with $t (-t)$.

For the case with uniform hoppings, the lattice in Fig.~1 has a four-site unit cell.  In
reciprocal space, with the reduced Brillouin zone
$(|k_x|, |k_y| \leq \pi/2, |k_z| \leq \pi)$,
the Hamiltonian can be written as
\begin{equation}\label{eq2}
H_0=\sum_{\bf{k}\sigma}
\psi_{\bf{k}\sigma}^{\dagger} {\cal H}_0(\bf{k})
\psi_{\bf{k}\sigma}^{\phantom{\dagger}}
\end{equation}
with $\psi_{\bf{k}\sigma}^{\phantom{\dagger}}=(c_{A\sigma},c_{B\sigma},c_{C\sigma},c_{D\sigma})^{T}$
and ${\cal H}_0({\bf k})$
\begin{equation}\label{eqk}
\left(
                    \begin{array}{cccc}
                      -2t\cos k_z & 0 & 2t\cos k_x & -2t\cos k_y \\
                       0 & -2t\cos k_z & 2t\cos k_y & 2t\cos k_x \\
                       2t\cos k_x& 2t\cos k_y & 2t\cos k_z & 0 \\
                       -2t\cos k_y& 2t\cos k_x & 0 & 2t\cos k_z \\
                    \end{array}
                  \right). \nonumber
\end{equation}
The above matrix can be written compactly as
\begin{equation}\label{eq3}
{\cal H}({\bf k})=-2t\cos k_z\tau_zI+2t\cos k_x\tau_xI+2t\cos k_y \tau_y\sigma_y
\end{equation}
with $\tau_{x,y,z}, \sigma_{y}$ the Pauli matrices and $I$ the $2\times 2$ identity matrix. The energy spectrum is given by
\begin{equation}\label{eq4}
E_{\bf k} = \pm 2t\sqrt{\cos^2 k_x+\cos^2 k_y+\cos^2 k_z},
\end{equation}
and the noninteracting system is a 3D semimetal with two inequivalent
Dirac points at ${\bf K}_{1,2}=(\pi/2,\pi/2,\pm \pi/2)$.

Manipulating $t_{lj}$ with the pattern shown in Fig.1, the unit cell is doubled along the z-direction. The two Dirac points are folded to the same point $(\pi/2,\pi/2,\pi/2)$ in the reduced Brillouin zone. The Hamiltonian in momentum space becomes,
\begin{equation}\label{eq5}
H=\sum_{\bf k\sigma}\Psi^{\dagger}_{{\bf k}\sigma}{\cal H}({\bf k})\Psi_{{\bf k}\sigma},
\end{equation}
with
\begin{eqnarray}\label{eq22}
{\cal H}({\bf k})=\left(
                    \begin{array}{cc}
                      h_{11} & h_{12} \\
                      h^{\dagger}_{12} & h_{22} \\
                    \end{array}
                  \right), \nonumber
\end{eqnarray}
and
\begin{eqnarray*}
h_{11}=\left(
  \begin{array}{cccc}
    0 & 0 & 2t_+(k_x) & -2t_-(k_y) \\
    0 & 0 & 2t_+(k_y) & 2t_-(k_x)  \\
    2t_+(k_x) & 2t_+(k_y) & 0 & 0 \\
    -2t_-(k_y) & 2t_-(k_x) &0  & 0
  \end{array}
\right),\\
h_{22}=\left(
  \begin{array}{cccc}
      0 & 0 & 2t_-(k_x) & -2t_-(k_y) \\
      0 & 0 & 2t_-(k_y) & 2t_-(k_x) \\
      2t_-(k_x) &2t_-(k_y) & 0 & 0 \\
       -2t_-(k_y) &2t_-(k_x) & 0 & 0
  \end{array}
\right),\\
h_{12}=\left(
  \begin{array}{cccccccc}
     0 & 0 & 0 & -2t_-(k_z) \\
     0 & 0 & -2t_-(k_z) & 0 \\
     0 & 2t_+(k_z) & 0 & 0 \\
     2t_-(k_z) & 0 & 0 & 0
  \end{array}
\right),
\end{eqnarray*}
where $t_{\pm}(k_i)=t_{\pm}\cos k_i$ ($i=x,y,z$).
The basis is
\begin{equation}\label{eq6a}
\Psi_{{\bf k}\sigma}=\{c_{A\sigma},c_{B\sigma},c_{C\sigma},c_{D\sigma},c_{A'\sigma},c_{B'\sigma},c_{C'\sigma},c_{D'\sigma}\}^{T}. \nonumber
\end{equation}
One easily obtains the energy spectrum,
\begin{equation}\label{eq6}
E_{\bf k}=\pm 2t_{\pm}\sqrt{\cos^2 k_x+\cos^2 k_y+\cos^2 k_z},
\end{equation}
which describes two-species 3D Dirac fermions with different velocities $2t_{\pm}$. In the above spectrum, the branch with $t_{+}$ is non-degenerate, and the one with $t_{-}$ is three-fold degenerate.
In the rest of the manuscript, we let $t_{\pm}=(1\pm\alpha)t$ and take $t_{+}=1$ as the energy scale for the birefringent case.

We further consider the Hubbard interaction,
\begin{equation}\label{eq7}
H_{U}=\sum_{i}
U(n_{i\uparrow}-\frac{1}{2})(n_{i\downarrow}-\frac{1}{2}).
\end{equation}
The total Hamiltonian $H=H_0+H_{U}$ can be solved numerically by means of the
DQMC method\cite{scalapino_1981,hirsch1983,hirsch1985,white1989}. In this approach, one decouples the on-site
interaction term through the introduction of an auxiliary
Hubbard-Stratonovich field (HSF).  The fermions are integrated
out analytically, and then the integral over the HSF is performed
stochastically.  The
only errors are those associated with the statistical sampling, the
finite spatial lattice
and inverse temperature discretization.
All
are well-controlled in the sense that they can be systematically reduced
as needed, and further eliminated by appropriate extrapolations. The
systems we studied have $N=L \times L \times L$ sites with $L$ up to
$10$. The temperatures accessed are down to $T/t\sim 0.1$. A Trotter discretization $\Delta \tau=0.1$ is used, which is small enough so that Trotter errors are comparable to the statistical uncertainty from the Monte Carlo sampling. Results represent averages of $10\sim 20$ independent runs with several hundreds sweeps each depending on the temperatures and the lattice sizes.

\section{Mean-field theory}

\begin{figure}[htbp]
\centering \includegraphics[width=7.5cm]{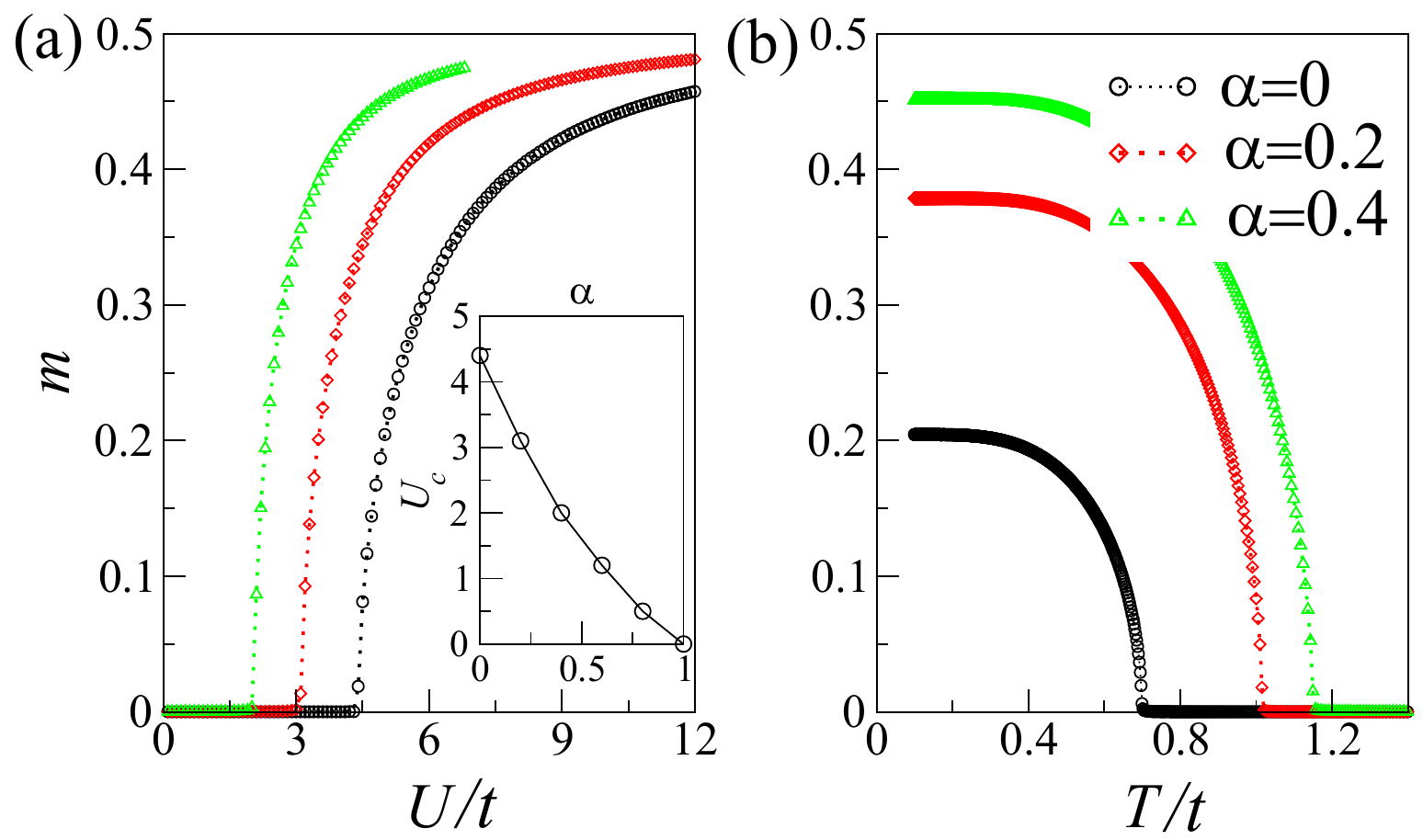} \caption{The mean-field parameter $m$ calculated self-consistently for several values of $\alpha$ as a function of: (a) interaction $U$ at zero temperature; (b) temperature $T$ at $U/t=10$. The critical interaction for the Dirac semimetal ($\alpha=0$) to AF insulator transition is about $U_c/t=4.3$ at $T=0$. Although the order parameters differ from each other in the unit cell for the birefringent cases with $\alpha\neq 0$, their transition points are the same, and we only show the order parameter on the $A$-site, i.e., $m_{1}$.}
\label{fig3}
\end{figure}

In the mean field approximation, the interaction is decoupled as,
\begin{equation}\label{eq8}
n_{i\uparrow}n_{i\downarrow}\approx \langle n_{i\downarrow}\rangle n_{i\uparrow}+\langle n_{i\uparrow}\rangle n_{i\downarrow}-\langle n_{i\downarrow}\rangle \langle n_{i\uparrow} \rangle.
\end{equation}
At half filling, we consider AF order, and write $\langle n_{i\uparrow}\rangle=\frac{1}{2}\pm m$, $\langle n_{i\downarrow}\rangle=\frac{1}{2}\mp m$ ($\pm$ depending on the sublattices). Then the four-fermion interaction term is decoupled as
\begin{eqnarray}\label{eq9}
\sum_{i} n_{i\uparrow}n_{i\downarrow}&\approx& \sum_{i\in A}(-m n_{i\uparrow}+m n_{i\downarrow}) \\ \nonumber
&+&\sum_{i\in B}(m n_{i\uparrow}-m n_{i\downarrow})+E_{0},
\end{eqnarray}
where the constant $E_0=\frac{1}{4}NU+NUm^2$.

In momentum space with the eight-component basis $\Psi_{{\bf k}\sigma}$, the following term is added to the non-interacting Hamiltonian Eq.(\ref{eq1}),
\begin{equation}\label{eq10}
H_{m}=\pm mU [\text{diag}(-1,-1,1,1,-1,-1,1,1) ],
\end{equation}
where $+(-)$ is for the spin-up (down) subsystem.
The energy spectrum of the total Hamiltonian becomes,
\begin{equation}\label{eq11}
E^{\pm}_{\bf k}=\pm \sqrt{4t^2(\cos^2 k_x+\cos^2 k_y+\cos^2 k_z)+(Um)^2},  \nonumber
\end{equation}
each of which is four-fold degenerate. The energy spectrum is the same for both spin subsystems.
$m$ is obtained by minimizing the free energy,
\begin{equation}\label{eq12}
F=-\frac{8}{\beta}\sum_{\bf k}\ln(1+e^{-\beta E^{\pm}_{\bf k}})+E_0,
\end{equation}
i.e., $\frac{\partial F}{\partial m}=0$, and the following self-consistent equation is obtained,
\begin{eqnarray}\label{eq13}
1=\frac{4U}{N}\sum_{\bf k}\frac{\tanh (\frac{\beta E^{+}_{\bf k}}{2})}{E^{+}_{\bf k}}.
\end{eqnarray}

For the case of birefringent Dirac fermions, the order parameters differ from each other in the unit cell. Suppose the order parameter on the $i$-th site is $m_i(i=1,2,...,8)$, then the term decoupled from the Hubbard interaction is,
\begin{equation}\label{eq10a}
H'_{m}=\pm U [\text{diag}(-m_{1},-m_{2},m_{3},m_{4},-m_{5},-m_{6},m_{7},m_{8}) ].
\end{equation}
The energy spectrum of the total Hamiltonian $H=H_{0}+H'_{m}$ does not have a simple analytical expression. With the energy eigenvalues $E_{{\bf k},\sigma}^{(j)}$ calculated numerically, the self-consistent equations for $m_i$ are as follows,
\begin{eqnarray}\label{eq13a}
m_i=-\frac{4}{NU}\sum_{ {\bf k},\sigma }\sum_{j=1}^{8}\frac{1}{1+e^{-\beta E_{{\bf k},\sigma}^{(j)}}} \frac{\partial E_{{\bf k},\sigma}^{(j)} }{\partial m_i}.
\end{eqnarray}

The determined order parameters are shown in Fig.\ref{fig3}. For zero temperature, $m$ changes to a non-zero value at the critical interaction $U_c$, marking the Dirac semimetal to AF insulator transition[see Fig.\ref{fig3}(a)]. $U_c$ is estimated to be about $\sim 4.3t$ for $\alpha=0$. Figure \ref{fig3}(b) shows the order parameter $m$ as a function of temperature. $m$ vanishes at a critical $T_c$, determining the N\'eel temperature. Quantum fluctuations usually modify the mean-field values, which will be clarified in the following DQMC simulations.

For birefringent Dirac fermions, the velocity $v_{F}=(1-\alpha)/(1+\alpha)$ of the outer Dirac cones can be continuously tuned by the ratio $\alpha$, while the inner cone velocity fixes the bandwidth. As shown in Fig.\ref{fig3}(a), the critical interaction strength is continuously decreased to zero with $\alpha$. In the $\alpha=1$ limit, the geometry is the perovskite lattice\cite{weeks_2010}, where the outer cones become exactly flat, and long-ranged AF order exists for all $U>0$. The setup provides an ideal system to study the effect of the velocity on the AF critical interaction, and verifies the velocity is the dominating parameter in determining the AF transition. In contrast, the N\'eel temperature increases as the velocity is decreased. It can be qualitatively understood that the AF order becomes more stabilized by decreasing the kinetic energy, and thus more robust to thermal quantum fluctuations.

\begin{figure}[htbp]
\centering \includegraphics[width=7.5cm]{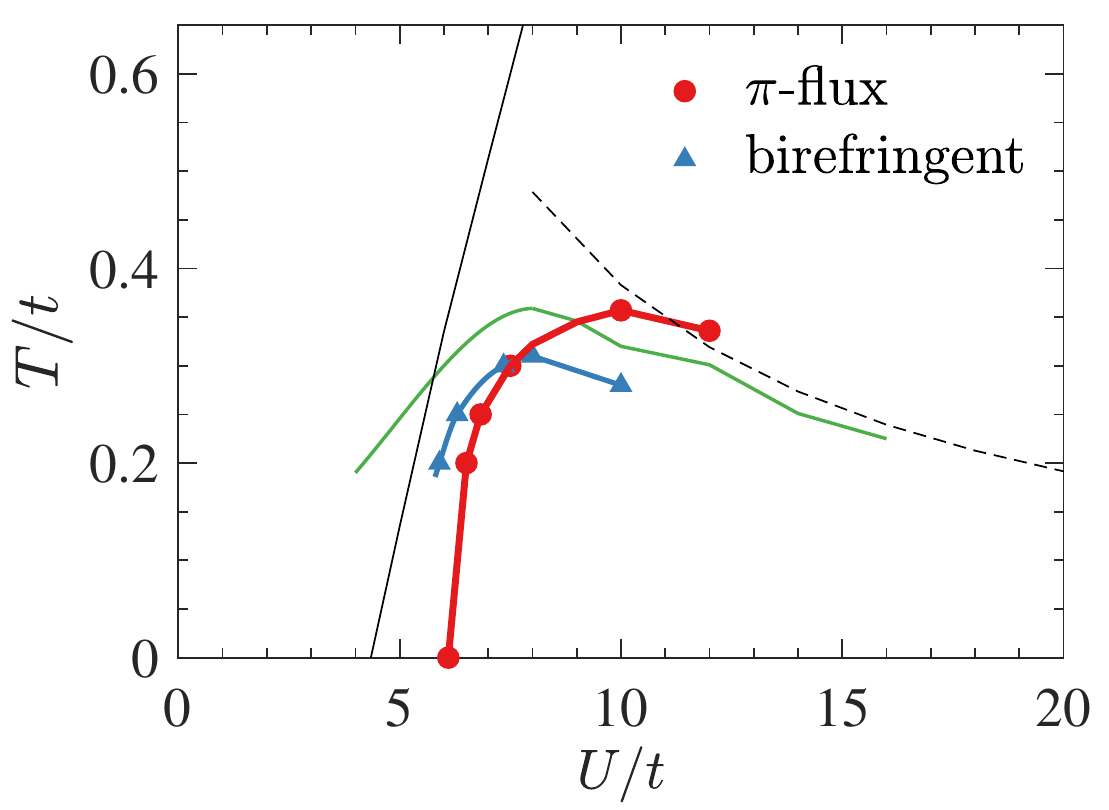} \caption{The phase diagram in the $(U,T)$ plane. The symbols represent DQMC data. The solid black curve is from self-consistent mean-field theory and the dashed black line is the strong coupling expression $T_N=3.83/U$. The green curve represents the AF phase boundary for the normal cubic lattice determined by DQMC and the numerical linked cluster expansion\cite{scalettar1989,ehsan2016, staudt2000}. We consider the anisotropy ratio $\alpha=0.1$ in the birefringent model.}
\label{fig4}
\end{figure}

\section{DQMC simulations}

\begin{figure}[htbp]
\centering \includegraphics[width=8.cm]{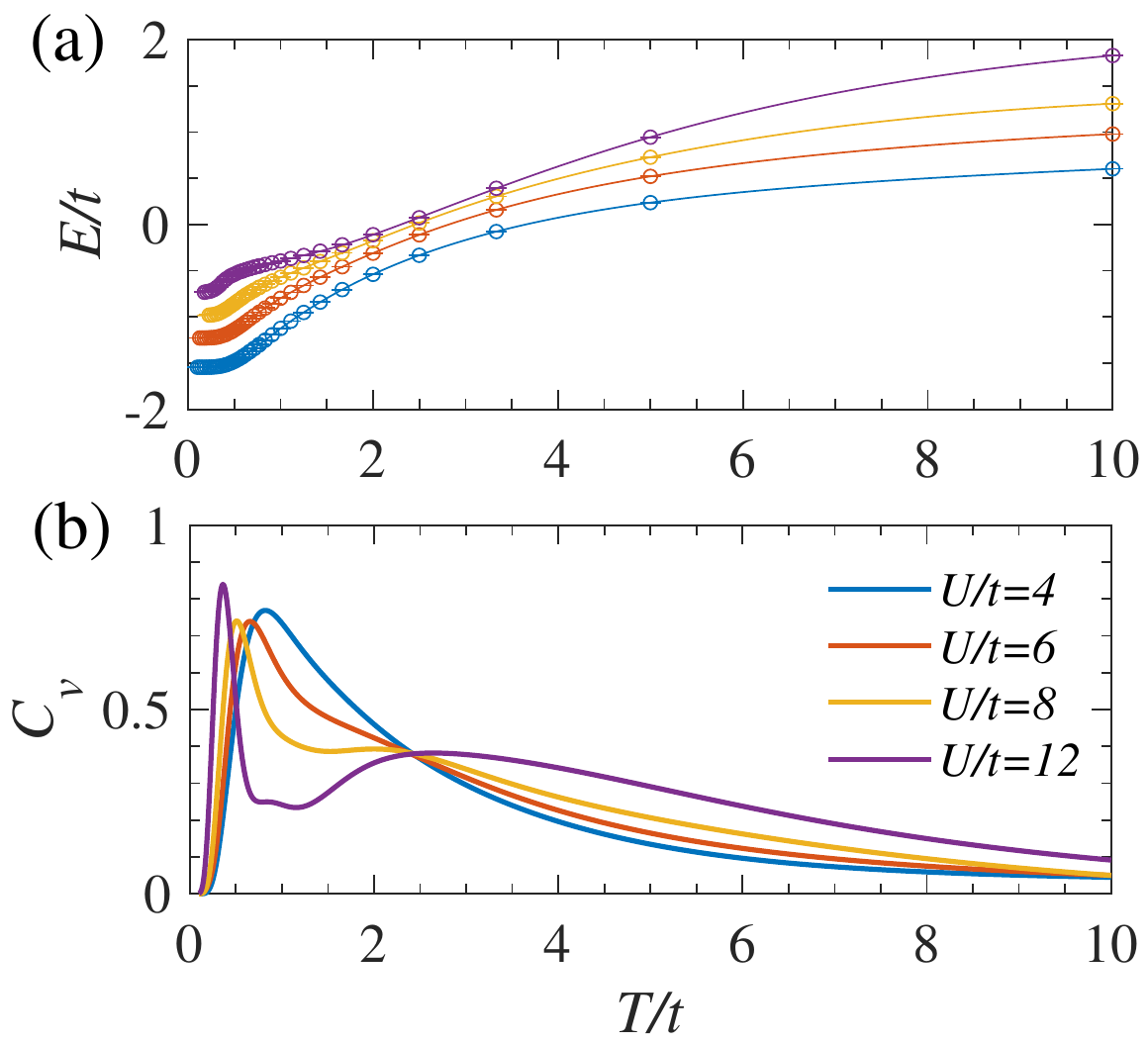} \caption{(a) The average energy per site vs temperature at half filling for several values of the interaction strength. Symbols are from DQMC simulations and curves are fitting functions. (b) Specific heat by directly differentiating the fitting function. The lattice has $N=L^3$ sites with $L=6$. The integer in the fitting function is $M=8$, which gives consistent results. Here $\alpha=0$ corresponding to one species of 3D Dirac fermion.}
\label{fig5}
\end{figure}

We first use DQMC to calculate the expectation value of the energy $E=\langle H\rangle$. To evaluate the specific heat at finite temperature,  the numerical data for $E(T_n)$ is matched to the functional form
\begin{eqnarray}\label{eq14}
E(T)=E(0)+\sum_{l=1}^{M}c_{l}e^{-\beta l\Delta},
\end{eqnarray}
where the parameters $c_{l},\Delta$ are found using the least squares method. The single functional form in Eq.(\ref{eq14}) has the correct low- and high-temperature limits $C(T)\rightarrow 0$\cite{paiva2001,paiva2013,paiva2015}. As shown in Fig.\ref{fig5}(a), $E$ is well fitted over a broad $T$ region. The specific heat is calculated by the standard formula,
\begin{eqnarray}\label{eq15}
C(T)=\frac{dE(T)}{dT}.
\end{eqnarray}
In Fig.\ref{fig5}(b), there is an evolution from one- to two- peak structures as $U$ is increased. In contrast, $C(T)$ always has a two-peak structure for both weak and strong couplings on the normal cubic lattice\cite{scalettar1989,ehsan2016}. The high-$T$ peak is a 'charge peak' at $T\sim U$, which corresponds to the suppression of the double occupancy and the decrease of the potential energy. The low-$T$ peak is due to the kinetic-energy decrease associated with the AF ordering, and thus the location is proportional to the exchange coupling, i.e., $T\sim J=\frac{4t^2}{U}$. The one-peak structure in the $\pi$-flux cubic lattice at small $U$ reflects the fact that the ground state of the weakly interacting system lacks AF order and remains in the semimetal phase. Such an evolution with $U$ thus reflects the underlying quantum phase transition from semimetal to AF order.

\begin{figure}[htbp]
\centering \includegraphics[width=6.5cm]{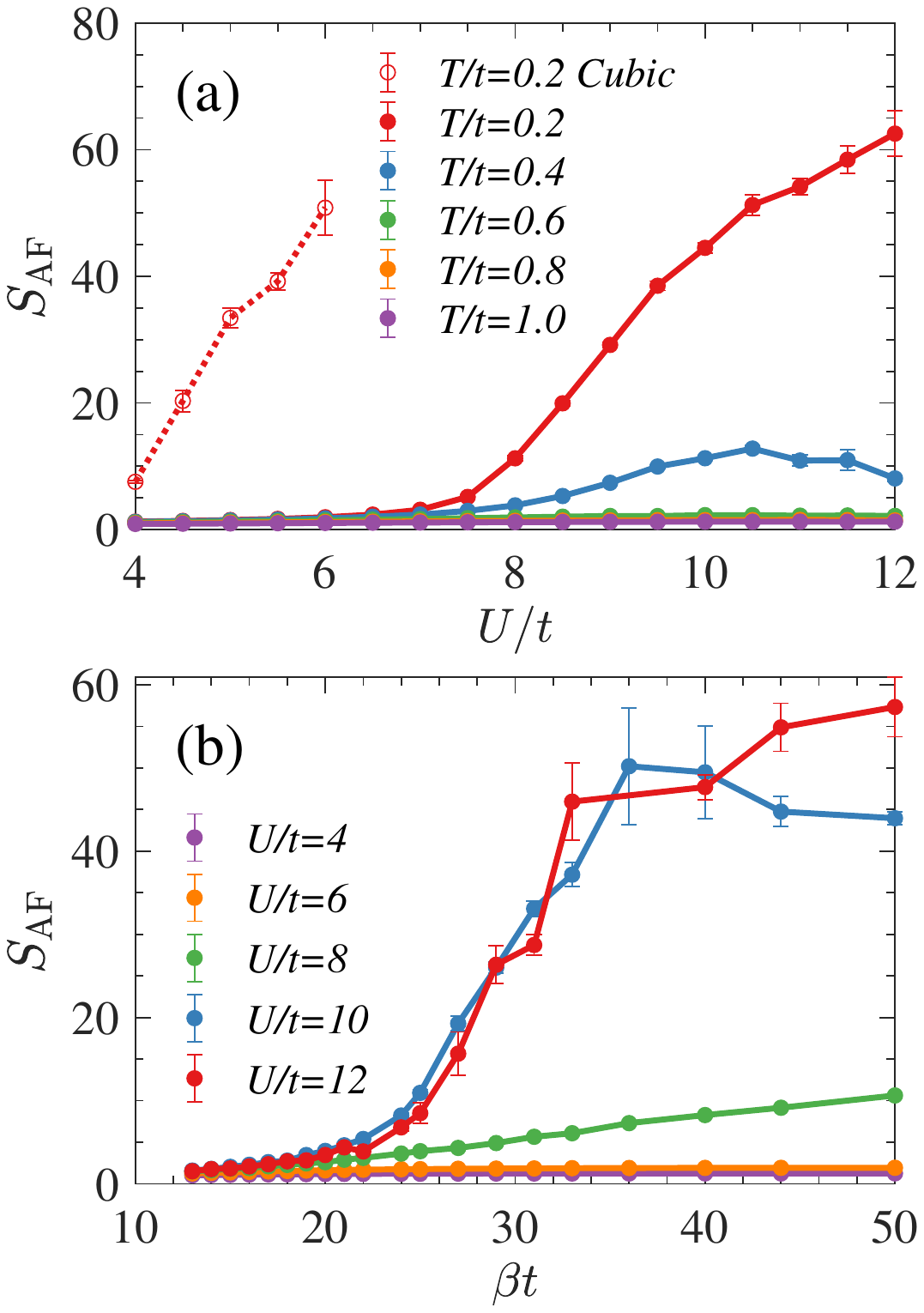} \caption{ (a) $S_{AF}$ vs interaction strength for several $T$. (b) $S_{AF}$ vs inverse temperature for several $U$. In (a), $S_{AF}$ at $T/t=0.2$ on the normal cubic lattice is plotted for comparison, where antiferromagnetic order develops in the ground state for arbitrary $U>0$. The lattice has $N=8^3$ sites. Here $\alpha=0$.}
\label{fig6}
\end{figure}

We then calculate the equal-time spin structure factor\cite{lin1987,hirsch1983,hirsch1985,hirsch1989,hirsch1987,white1989},
\begin{eqnarray}\label{eq15a}
S_{AF}({\bf Q})=\frac{1}{N}\sum_{i,j}e^{i{\bf Q}\cdot ({\bf r}_j-{\bf r}_i)}\langle {\bf S}_i\cdot {\bf S}_j\rangle,
\end{eqnarray}
where ${\bf Q}=(\pi,\pi,\pi)$ is the AF wave vector.
The results on a finite lattice with $N=8^3$ sites are shown in Fig.\ref{fig6}, where the temperature simulated reaches as low as $T/t=0.2$. Figure \ref{fig6} (a) shows the spin structure factor at fixed temperatures as a function of $U$. At high temperatures, $S_{AF}({\bf Q})$ remains almost zero for $U$ up to $12t$, suggesting that no long-range order exists and the system is in the paramagnetic phase above the N\'eel temperature.
Below the N\'eel temperature, there is a clear transition happening at finite $U$, from which the spin structure factor begins to grow significantly. For the temperature $T/t=0.4$, which is close to the N\'eel one at $U/t=10$, there is a peak at $U\sim 10t$, corresponding to the highest N$\acute{e}$el temperature. We also plot $S_{AF}({\bf Q})$ for the normal cubic lattice at $T/t=0.2$, whose curve is significant shifted toward the weakly-interacting region\cite{scalettar1989,paiva2011,ehsan2016}. This is expected, since the usual half-filled Hubbard model on the simple cubic lattice has a perfectly nested Fermi surface and thus an AF instability for infinitesimal $U$ \cite{scalapino_1986}. Thus by contrast with the normal cubic lattice, large enough interactions are needed to drive the AF transition for 3D Dirac fermions.

In Fig.\ref{fig6}(b), we show $S_{AF}({\bf Q})$ as a function of inverse temperature for fixed interaction strength. There is no change with decreasing temperature for weak interactions, consistent with a finite critical interaction. For large $U$, $S_{\mathrm{AF}}({\bf Q})$ starts to increase above $\beta t\sim 20$, and then saturates at finite values, indicating AF order develops below a critical temperature.

\begin{figure}[htbp]
\centering \includegraphics[width=8.5cm]{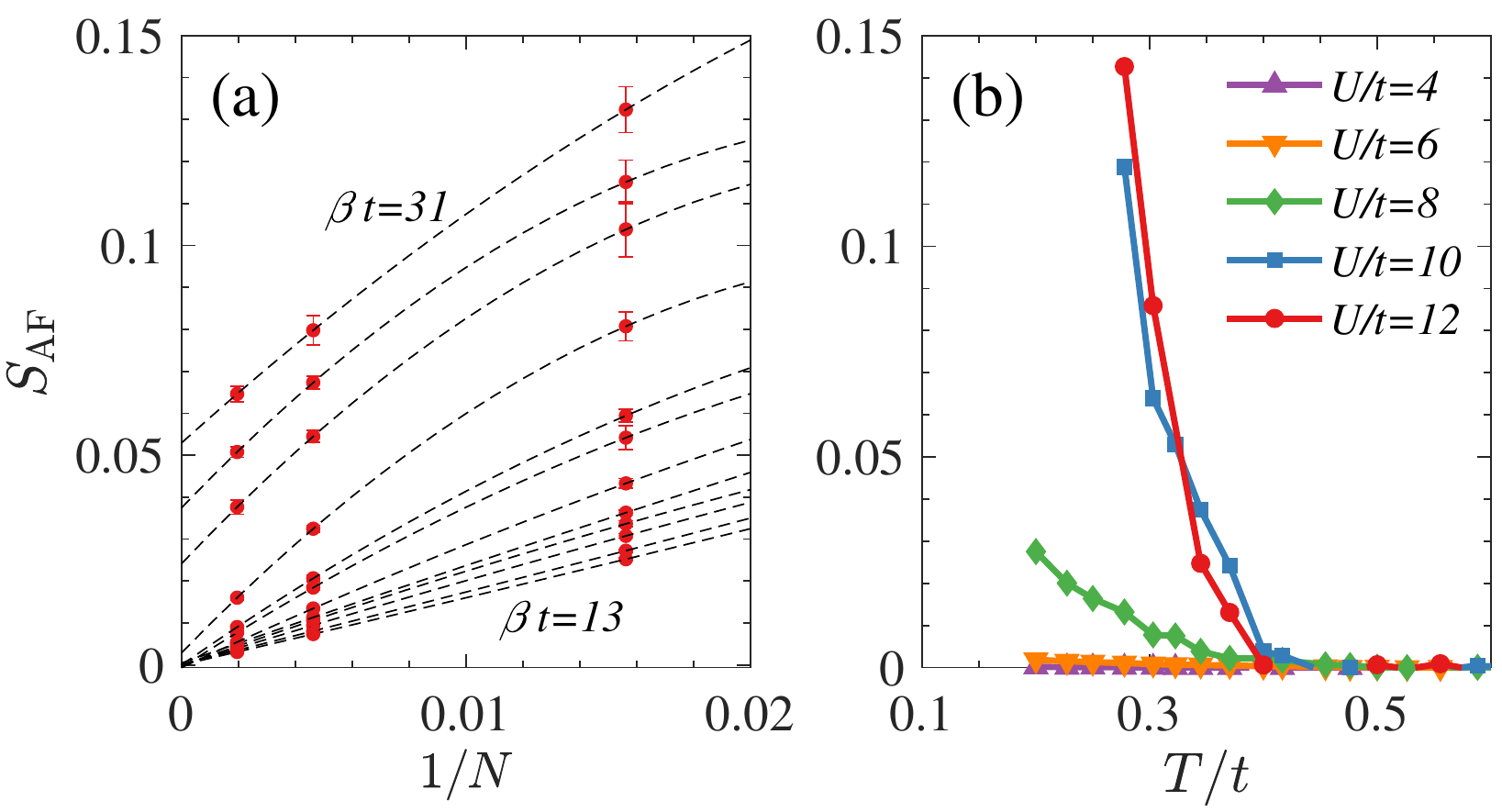} \caption{(a) Extrapolation of the spin structure factor at various temperatures for $U/t=10$. (b) Values of $S_{AF}/N$ extrapolated to the thermodynamic limit vs temperature for several $U$. Here $\alpha=0$.}
\label{fig7}
\end{figure}

The finite-size DQMC data should be extrapolated to the thermodynamic limit, and generally we have $S_{\mathrm{AF}}({\bf q})/N=S_0+f(L)$ with $f(L)\rightarrow 0$ for $L\rightarrow \infty$.  In the large-$U$ limit, the Hubbard model maps onto the spin$-1/2$ Heisenberg model on the cubic lattice. This model exhibits a finite-temperature AF transition in the 3D Heisenberg universality class, for which we expect $S_{AF}({\bf Q})/N$ to scale with the system size as $L^{-1-\eta}$ at the critical temperature, with $\eta=0.0375$ the order parameter anomalous dimension\cite{Zinn-Justin2010}. However in previous studies of the 3D Heisenberg model, only the $L\geq 10$ data show the expected scaling\cite{sandvik1998}, which are difficult for the DQMC method to access. The spin-wave theory predicts that the spin-spin
correlation function varies as the inverse of the distance, and the leading correction is $\sim 1/L$\cite{huse1988,sandvik1997,scalettar2009}. However with the limited lattice sizes, we find the data is best fit by a quadratic polynomial in $1/N$\cite{staudt2000}, i.e., $f(N)=\alpha/N+\beta/N^2+...$, with $\alpha, \beta$ the fitting coefficients[see Fig.\ref{fig7}(a)]. While the extrapolated values in Fig.\ref{fig7}(b) clearly demonstrate the existence of the critical interactions and temperatures, the critical values can only be qualitatively estimated. $T_c$ first increases as $U$ is enlarged. Then the curve of $T_c$ for $U/t=12$ becomes below that for $U/t=10$ near the transition, indicating the decrease of the critical temperature thereafter. Such a behavior is consistent with a dome region of the AF order in the phase diagram.

\begin{figure}[htbp]
\centering \includegraphics[width=8.5cm]{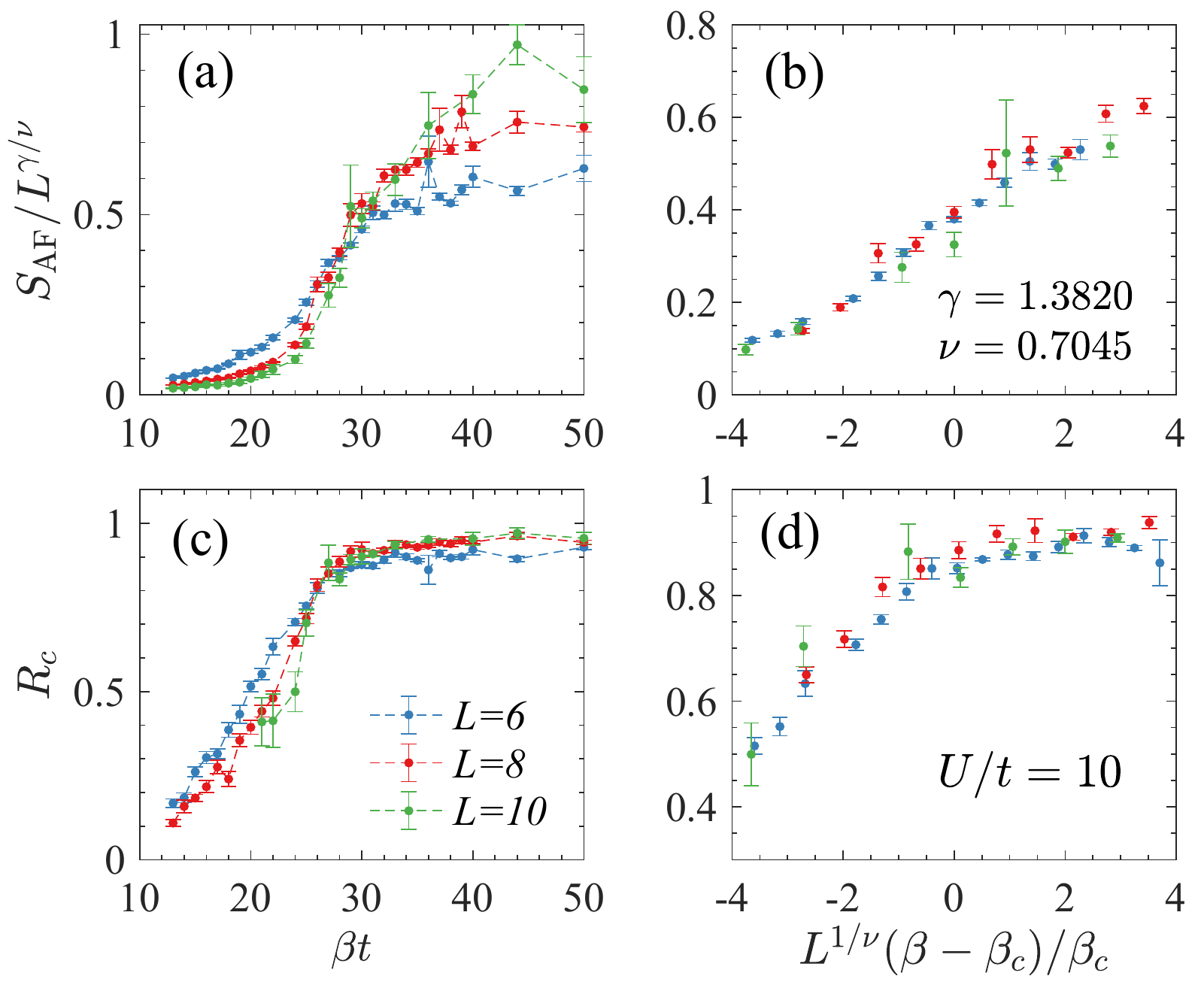} \caption{(a) Scaled spin structure factor $S_{\mathrm{AF}}/L^{\gamma/ \nu}$ as a function of $\beta$. (b) Best scaling collapse of $S_{\mathrm{AF}}/L^{\gamma/ \nu}$ gives  $T_{c}/t=0.36$. (c) Correlation ratio $R_c$ as a function of $\beta$. (d) Best scaling collapse of the correlation ratio, also giving the critical temperature $T_{c}/t=0.36$.  The scaling exponents are taken to be their values in the 3D Heisenberg universality class [see the inset of (b)], and provide a good universal crossing near the critical temperature in (a) and (c). Here $\alpha=0$.}
\label{fig8}
\end{figure}

The critical values can be determined more precisely by the crossing of the scaled $S_{\mathrm{AF}}(\mathbf{Q})$ with the universal critical exponents. The finite-temperature transition belongs to the Wilson-Fisher $O(3)$ universality class, describing the ferromagnetic transition in the 3D classical Heisenberg model\cite{stephanov1995,landau1991}. The critical exponents of the $O(3)$ model from Borel summation of the $\varepsilon$-expansion gives $\gamma=1.3820$ and $\nu=0.7045$ \cite{Zinn-Justin2010}. The best data collapse occurs at $T_c/t=0.36$ for $U/t=10$, which is consistent with the crossing of $S_{\mathrm{AF}}(\mathbf{Q})/L^{\gamma/ \nu}$ in Fig.\ref{fig8}(a).  We further verify the critical values from the crossings of invariant correlation ratio\cite{binder1981,lang2016,chenchuang2019,zhangyuxi2019},
\begin{eqnarray}\label{eq16a}
R_{c} \equiv 1-\frac{S_{\mathrm{AF}}(\mathbf{Q}+\delta \mathbf{q})}{S_{\mathrm{AF}}(\mathbf{Q})},
\end{eqnarray}
where $\delta \mathbf{q}$ points to a nearest-neighbor momentum in the Brillouin zone. In the presence (absence) of long-range order, we have $S_{\mathrm{AF}}(\mathbf{Q}+\delta \mathbf{q})\rightarrow 0 (S_{\mathrm{AF}}(\mathbf{Q}))$, and thus $R_c\rightarrow 1 (0)$. At the critical point, the use of $R_c$ is advantageous as it has smaller scaling corrections than $S_{\mathrm{AF}}(\mathbf{Q})$ itself. Since $R_c$ has no scaling dimension, we collapse the data with the scaling form $\mathcal{F}_{R_{c}}\left[L^{1 / \nu}\left(U-U_{c}\right) / U_{c}\right]$. As shown in Fig.\ref{fig8}(d), the best scaling collapse on the interval $[-4,4]$ gives the critical temperature $T_{c}/t=0.36$, consistent with the previous determined one. It is noted that $T_c$ obtained by the universal scaling qualitatively matches that obtained by finite-size extrapolation.

A similar analysis for other values of $U$ yields the phase boundary in Fig.\ref{fig4}. The AF dome shifts to the right in contrast to the normal cubic lattice. $T_c$ exhibits a maximum at $U/t\sim10$, reflecting the competition between the growth of the local moment and a reduction of the AF coupling with $U$. The local moment $m_z^2=\langle (n_{\uparrow}-n_{\downarrow})^2\rangle=1-2\langle n_{\uparrow}n_{\downarrow}\rangle$ at half filling. The double occupancy is suppressed by the interaction, resulting in the growth of $m_z^2$ and thus the AF order parameter $S_{AF}(\mathrm{Q})$ with $U$. While such a behavior can be understood in terms of the mean-field theory, the virtual transitions between spin-up and -down states by a second order hopping process are omitted therein, causing $T_c$ to keep growing artificially in the strong-coupling region. In fact, the AF coupling, which is $\sim t^2/U$, becomes dominant in determining the reduction of $T_c$ at large $U$.

\begin{figure}[htbp]
\centering \includegraphics[width=8.5cm]{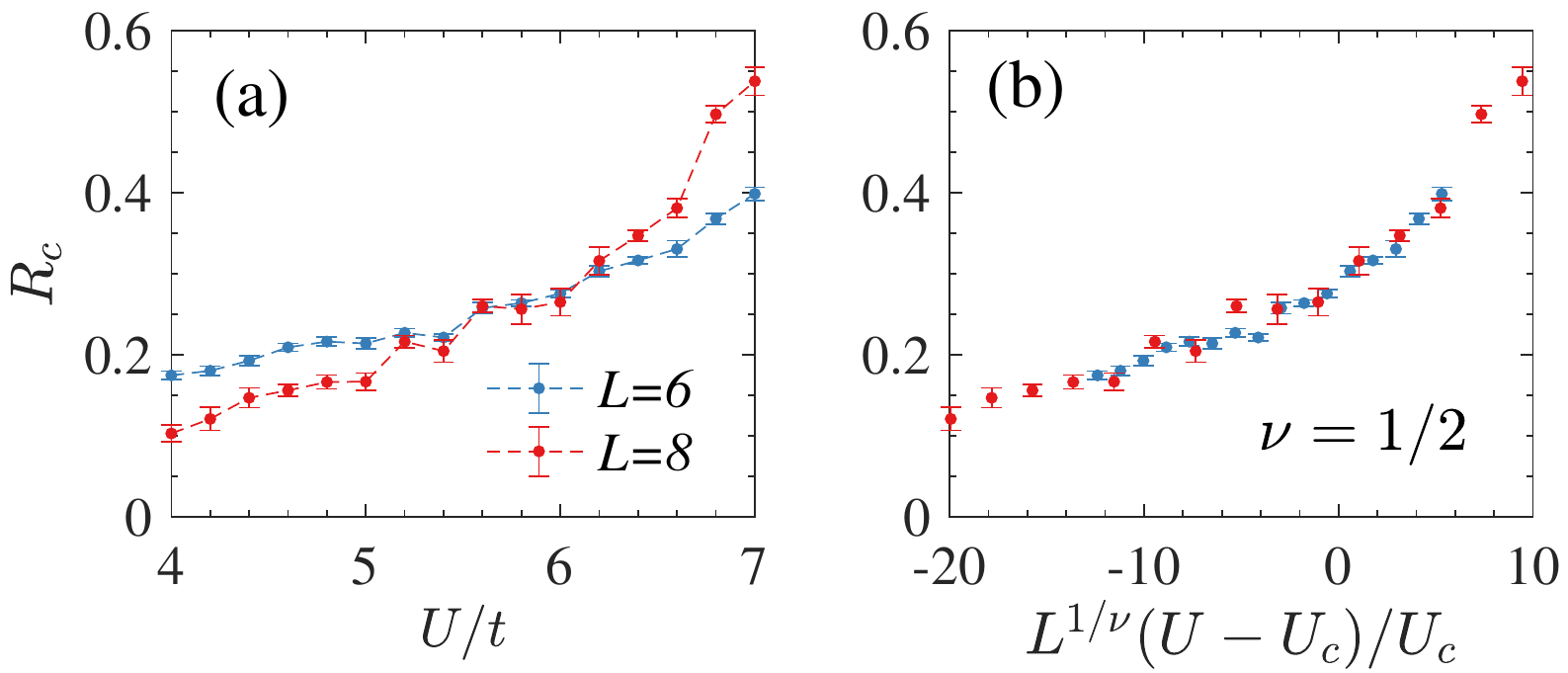} \caption{(a) Correlation ratio $R_c$ as a function of $U$. The critical value is $U_c/t\approx 6.1$, estimated from the intersection of $R_c$. (b) Scaling collapse of $R_c$ using the mean-field critical exponent $\nu=1/2$. Here the inverse temperature is $\beta t=L$. Here $\alpha=0$.}
\label{fig9}
\end{figure}

Besides the N$\acute{e}$el temperatures of thermal phase transitions, by contrast with both the Heisenberg model and the Hubbard model on the normal cubic lattice, there is also a quantum critical point at $U=U_c$ in the ground state, which is described by the Gross-Neveu universality class in (3+1)d. The correlation length exponent is the mean-field one, i.e., $\nu=1/2$. Since the dynamic critical exponent $z=1$ due to the emergent Lorentz symmetry, we take $\beta t=L$ to see the quantum critical scaling. The critical interaction is estimated from the intersection in Fig.\ref{fig9}(a), which is $U_c/t=6.1$. The mean-field critical exponent provides a good universal scaling collapse of the correlation ratio, further verifying the nature of the quantum phase transition[see Fig.\ref{fig9}(b)].

\begin{figure}[htbp]
\centering \includegraphics[width=7.cm]{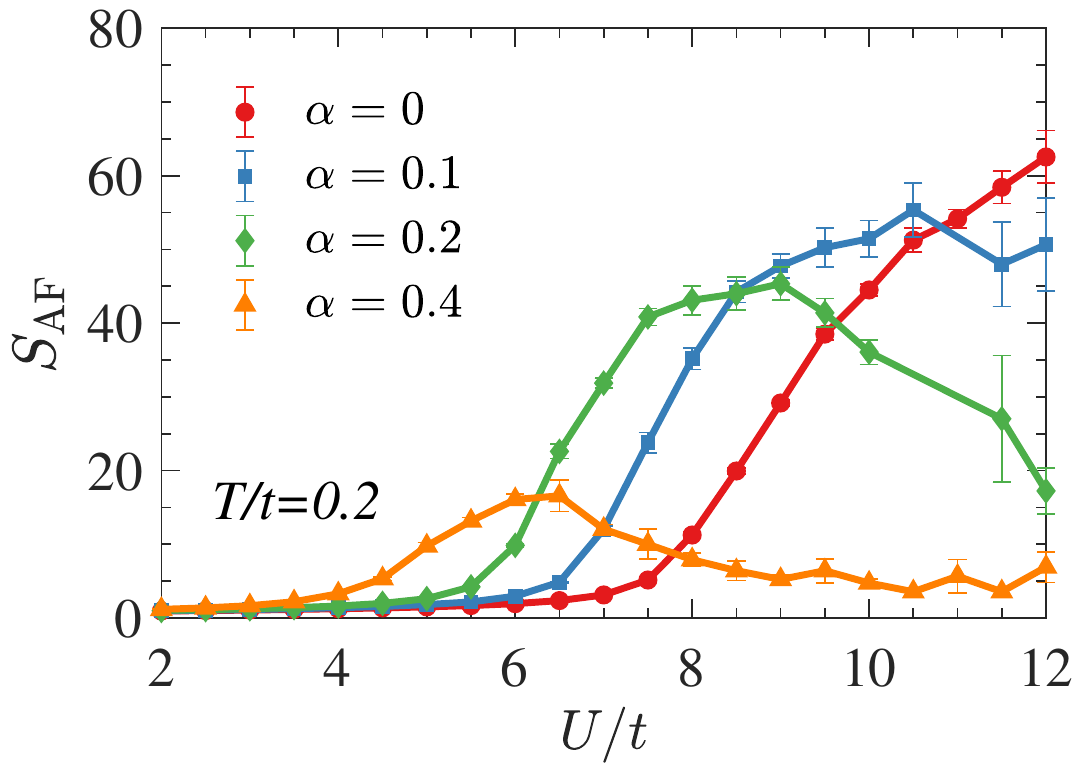} \caption{Spin structure factor vs interaction strength for birefringent Dirac fermions on a $L=8$ lattice. The parameter $\alpha$ continuously tunes the velocity of one species of Dirac fermions $v_{F}=(1-\alpha)/(1+\alpha)$ while the other one is fixed to unity. The temperature is $T/t=0.2$.}
\label{fig10}
\end{figure}

Finally we turn to the case of 3D birefringent Dirac fermions. Figure \ref{fig10} shows $S_{AF}$ on a $N=8^3$ lattice for different $\alpha$ as a function $U$ at $T/t=0.2$. At the $\pi$-flux point ($\alpha=0$), we have obtained the critical interaction $U_c/t=6.51$ for the semimetal-AF insulator transition at $T/t=0.2$. The curves of $S_{AF}$ are significantly shifted to the left as we increase $\alpha$, which suggests a decrease of $U_c$ with $\alpha$ and is consistent with the mean-field analysis. Meanwhile, the maximum value of $S_{AF}$ drops quickly, and the one at $\alpha=0.4$ already becomes about $\sim 1/3$ of that at $\alpha=0$. It implies the temperature $T/t=0.2$ is much closer to $T_{N}$ for large $\alpha$, and thus the N\'eel temperature decreases with increasing $\alpha$. We determine the phase boundary for the AF order at $\alpha=0.1$. As shown in Fig.\ref{fig4}, the dome moves to the left and its peak goes down, consistent with the above observations. Intuitively, the decrease of the N\'eel temperature might be associated with the exchange couplings on the modulated bonds, being increasingly weakened, and indeed are finally completely depleted from the lattice at $\alpha=1$.

\section{Conclusions}
The interaction-driven AF transitions of 3D Dirac fermions are investigated based on the $\pi$-flux model on a cubic lattice using DQMC simulations. We find the AF order only occurs above a finite critical interaction. While the thermal phase transitions belong to the 3D Heisenberg universality class, the critical behavior for the quantum critical point is consistent with the (3+1)d Gross-Neveu universality. The critical interaction strength and temperatures are evaluated by finite-size scaling of the spin structure factor, and the phase diagram in the $(U,T)$ plane is mapped. It is found that while the critical interaction of the AF transition changes from $U_c/t=0$ of the normal cubic lattice to $U_c/t=6.1$ for the $\pi$-flux lattice, the interaction where $T_N$ is largest here only becomes a bit larger than that of the cubic lattice.
We further study correlation effects in a birefringent Dirac fermion system, and quantify the effect of the velocity on the critical interaction strength.

Our findings unambiguously uncover correlation effects in 3D Dirac fermions. The 3D Hubbard model has been readily emulated using ultracold atoms in an optical lattice\cite{hart_2015}. There have been several successful methods to generate strong artificial magnetic fields, such as Raman assisted hoppings, and rotating the gases\cite{bloch_2011,dalibard_2011}. It is very possible the strong magnetic field along the $(1,1,1)$ direction needed for the $\pi$-flux model could be synthesized based on the cubic optical lattice. With state-of-art measurement techniques, our results may be verified experimentally.

\section*{Acknowledgments}
The authors thank E. Khatami, G. Batrouni and Wenan Guo for helpful discussions. H.G. acknowledges support from the NSFC grant No.~11774019, the Fundamental Research
Funds for the Central Universities and the HPC resources
at Beihang University.
Y.H. and S.F. are supported by the National Key Research and Development Program of China under Grant No. 2016YFA0300304, and NSFC under Grant Nos. 11974051 and 11734002. J.M. was supported by NSERC Grants No. RGPIN-2020-06999 and RGPAS-2020-00064, the CRC Program, NFRF, and the University of Alberta. R.T.S.~was supported by DOE grant DE-SC0014671.

\bibliography{ddirac}

\end{document}